\documentclass[aps,twocolumn,showpacs]{revtex4}
\usepackage[dvips]{graphics}

\begin{document}

\title{Metal-insulator transition in an aperiodic ladder network: an exact
result}

\author{Shreekantha Sil$^1$, Santanu K. Maiti$^2$ and Arunava Chakrabarti$^3$}

\affiliation{$^1$Department of Physics, Visva-Bharati, Santiniketan, West 
Bengal-731 235, India. \\
$^2$Department of Physics, Narasinha Dutt College, 129 Belilious Road, 
Howrah-711 101, India. \\
$^3$Department of Physics, University of Kalyani, Kalyani, West 
Bengal-741 235, India.} 

\begin{abstract}
We show, in a completely analytical way, that a tight binding ladder network 
composed of atomic sites with on-site potentials distributed according to the 
quasiperiodic Aubry model can exhibit a metal-insulator transition at multiple 
values of the Fermi energy. For specific values of the first and second 
neighbor electron hopping, the result is obtained exactly. With a more 
general model, we calculate the two-terminal conductance numerically. The 
numerical results corroborate the analytical findings and yield a richer 
variety of spectrum showing multiple mobility edges.
\end{abstract}


\maketitle 

\noindent
The classic problem of electron localization in low dimensional quantum 
systems has remained alive over the last fifty years, since its proposition 
in $1958$ by Anderson~\cite{ander58}. It is now well known that in one 
dimension even for arbitrarily weak disorder (almost) all the one-electron 
states are exponentially localized~\cite{ander58,lee85} and one never 
encounters mobility edges, that is, energy eigenvalues separating localized 
(insulating) states from the extended (conducting) ones. Last couple of 
decades however, have witnessed an interesting twist in the canonical wisdom 
about localization problem through the advent of the so called {\em correlated 
disordered} models in one dimension. It is now established that even in a one 
dimensional chain of atomic sites, short range~\cite{dun90,san94} or long
range~\cite{fabf98,domin03} positional correlations  
between the constituents can induce unscattered (extended) wave functions 
at specific values of the electron-energy~\cite{dun90,san94} or even a 
crossover from localized to extended states~\cite{fabf98,domin03}. 

The search for a mobility edge in one dimensional lattices has always been 
an intriguing problem that started earlier with certain aperiodic lattices 
which bridge the gap between a perfectly periodic lattice and a completely 
random one. A famous example is the single band Aubry-Andre (AA) 
model~\cite{aubry79} in which the on-site potential in a $1$-d chain of 
lattice constant $a$ is described deterministically by, 
$\epsilon_n=\lambda \cos(Qna)$ with $Q$ being an irrational multiple of 
$\pi$. 
It is to be appreciated that an AA-model represents a certain class of 
almost periodic lattices which are much different from conventional 
randomly disordered systems, and displays a special kind of `order', 
called the quasiperiodic order. The system however
lacks translational periodicity. 
In a one dimensional chain with nearest neighbor hopping integral 
$t$ for example, and with such a potential (also called a Harper potential) 
the amplitude $f_n$ of the wave function at the $n$th site of the lattice 
can be obtained from the eigenvalue equation 
\begin{equation}
\left[E-\lambda \cos(Qna)\right]f_n=t(f_{n+1} + f_{n-1})
\label{equ2}
\end{equation}
In the AA-model, with $\lambda > 2t$ all the single particle states are 
exponentially localized with Lyapunov exponent $ln(\lambda/{2t})$, while 
$\lambda < 2t$ makes the states (all of them) extended~\cite{aubry79,soko85}, 
thus exhibiting a metal-insulator transition in parameter space. No 
mobility edges exist for this model. Another premier feature of this 
model is its self duality, which means that if one makes 
the transformation~\cite{soko85}, 
\begin{equation}
f_n = \sum_{m=-\infty}^{\infty} g_m e^{imnQa},~~~
g_m = \sum_{n=-\infty}^{\infty} f_n e^{-imnQa} 
\label{equ3}
\end{equation}
then the coefficients $g_m$ satisfy 
\begin{equation}
\left[E-2t \cos(Qma)\right]g_m 
= \frac{\lambda}{2} (g_{m+1} + g_{m-1})
\end{equation}
The $g_m$-equation is exactly of the same form as Eq.~(\ref{equ2}) with the 
roles of $t$ and $\lambda$ interchanged. It can be shown~\cite{soko85} that 
if the eigenstates given by $g_m$ are localized in reciprocal space, 
then the eigenstates of Eq.~(\ref{equ2}) will be extended in real space, 
and vice versa. Incidentally, several variants of the Aubry model have 
also been examined to detect the signature of mobility edges even in 
$1$-d~\cite{eco82,rolf90,andrej01}.

\begin{figure}[ht]
{\centering\resizebox*{8cm}{3cm}{\includegraphics{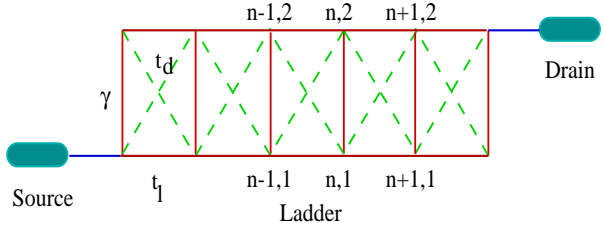}} \par}
\caption{Schematic view of a ladder attached to two electrodes.}
\label{ladder}
\end{figure}

In this letter we investigate the electronic spectrum of an AA-ladder 
network built by fixing two identical AA-chains laterally 
(see Fig.~\ref{ladder}). The motivation behind the present work is twofold. 
First, we wish to investigate if the interplay of the quasi-one dimensional 
structure of the network and the AA duality, which is still preserved, leads 
to any possibility of a metal-insulator (MI) transition even within the 
standard 
form of the Harper potential. If it is true, then a ladder network such as 
this, could be used as a switching device, the design of which is of great 
concern in the current era of nanofabrication. Interestingly, research in 
AA-models in 1-d and its variations have been 
rekindled recently in the context of potential design of aperiodic optical 
lattices~\cite{drese97,sankar06}. Therefore, the question of the existence 
of MI transition in a system with `pure' AA (Harper) potential can be 
addressed with a renewed interest. Secondly, the ladder networks have 
recently become extremely important in the context of understanding the 
charge transport in double stranded DNA~\cite{macia06,rudo07}. Experimental
results on DNA transport report wide range of behavior, from almost 
insulating~\cite{pab}, semiconducting~\cite{por} to even metallic~\cite{rak}, 
that can be attributed to many experimental complications, such as the
preparation of the sample, sample-electrode contact etc. In addition to
this, Mrevlishvili~\cite{geor} experimentally observed oscillations in
the specific heat of DNA structures at low temperatures, results that
have been subsequently explained by Moreira {\em et al.}~\cite{more}
considering a quasiperiodic sequence of the nucleotides. The results 
of reference~\cite{more}, compare remarkably well with numerical results
obtained for $Ch22$ human chromosomes. It is to be appreciated that the helical
structure of the double stranded DNA is expected to affect the periodicity
of the effective site potentials on the ladder, and introduce incommensurate
periods in the system. In view of 
this, the examination of the electronic spectrum of a ladder network 
comprising of aperiodically varying site potentials might throw new light 
into the behavior of electrons, both in the context of basic physics and 
possible technological applications including DNA devices.

We adopt a tight binding formalism, incorporate nearest and next nearest 
neighbor hopping inside a plaquette of the ladder and show that such a 
system exhibits a re-entrant MI transition. Most interestingly, for a 
selected set of the Hamiltonian parameters we are able to provide an 
exact analysis which shows the existence of mobility edges. We begin by 
referring to Fig.~\ref{ladder}. The Hamiltonian of the ladder network 
is given by, 
\begin{equation}
{\mathbf H}=\sum_n \bf{\epsilon_n c_n^{\dagger} c_n} + 
\bf{t \sum_n c_n^{\dagger} c_{n+1}} + h.c.
\end{equation}
where 
\begin{equation}
{\bf c_n}=\left(\begin{array}{c} c_{n,1} \\ c_{n,2} \end{array}\right),~
\bf{\epsilon_n}  =  \left( \begin{array}{cc} \epsilon_{n,1} & \gamma \\
\gamma & \epsilon_{n,2} \end{array} \right), ~
\bf{t}  =  \left( \begin{array}{cc} t_l & t_d \\
t_d & t_l \end{array} \right) \nonumber \\
\label{equ6}
\end{equation}
In the above, $c_{n,j}$ ($c_{n,j}^{\dagger}$) are the annihilation (creation)
operator at the $n$th site of the $j$th ladder, 
$\epsilon_{n,1}=\epsilon_{n,2}=\lambda \cos(Qna)$ is the on-site 
potential at the $n$th site of the $j$th ladder, $\gamma$ is the vertical 
hopping between the $n$th sites of the two ladders, $t_l$ is the nearest 
neighbor hopping integral between the $n$th and the ($n+1$)th sites of 
every arm and $t_d$ is the next nearest neighbor hopping within a plaquette 
of the ladder (see Fig.~\ref{ladder}).

We describe the system in a basis defined by the vector
\begin{equation}
{\bf f_n} = \left(\begin{array}{c} f_{n,1} \\ f_{n,2} \end{array}\right)
\label{equ8}
\end{equation}
where, $f_{n,j}$ is the amplitude of the wave function at the $n$th site of 
the $j$th arm of the ladder, $j$ being equal to $1$ or $2$. Using this basis, 
our task boils down to obtain solutions of the difference equation
\begin{equation}
(E{\bf I}-{\bf \epsilon_n})  {\mathbf f_n} = {\bf t}({\bf f_{n+1}} + 
{\bf f_{n-1}})
\label{equ9}
\end{equation}
At first, we proceed to show the existence of multiple mobility edges in 
such a system in an analytically exact way. For this, we choose $t_d=t_l$, 
and make the following transformation to the reciprocal space for each arm
of the ladder and arrive at a difference equation in the reciprocal space, 
\begin{eqnarray}
\left[\left\{E-2t_l \cos(Qma)\right\}{\bf I} \right . &-& \left .
\left\{2t_l \cos(Qma)+\gamma \right\}{\bf \sigma_x} \right]{\bf g_{m,j}} 
\nonumber \\
&=& (\lambda/2) {\bf I} \left({\bf g_{m+1,j}} + {\bf g_{m-1,j}} \right)  
\label{equ11}
\end{eqnarray}
where, {$\bf \sigma_x$} is the usual Pauli matrix. We now diagonalize the 
{$\bf \sigma_x$} matrix by a similarity transformation using a matrix 
$\mathbf S$, and define $|{\phi_m}\rangle = {\bf S} |{g_m}\rangle$
The difference equation (\ref{equ11}) now decouples into,  
\begin{equation}
(E + \gamma) \phi_{m,2} = (\lambda/2) (\phi_{m+1,2} + \phi_{m-1,2} )
\label{equ13}
\end{equation}
\begin{equation}
\left[E-\gamma-\Delta \cos(Qma)\right] \phi_{m,1} = (\lambda/2) 
(\phi_{m+1,1} + \phi_{m-1,1})
\label{equ14}
\end{equation}
Here, $\phi_{m,1}$ and $\phi_{m,2}$ are the elements of the column vector 
${\phi_m}$ and $\Delta=4t_l$. It is interesting to observe that, the 
Eq.~(\ref{equ13}) above corresponds to a perfectly ordered chain with 
nearest neighbor hopping integral equal to $\lambda/2$ in the reciprocal 
space. This implies that, for $-\lambda-\gamma < E < \lambda -\gamma$ we 
have a gap less continuous spectrum in the reciprocal space. 
Eq.~(\ref{equ14}) on the other hand, represents the familiar single band 
Aubry model for which all states are localized or extended if 
$\Delta > \lambda$, or, $\Delta < \lambda$ respectively.
We can now extract information about the nature of eigenfunctions by 
considering the two Eqs.~(\ref{equ13}) and (\ref{equ14}) simultaneously. 
\vskip 0.05in
\noindent
{\em Case I:} $|E+\gamma| < \lambda$ and $\Delta > \lambda$
\vskip 0.05in
We focus on the pair of Eqs.~(\ref{equ13}) and (\ref{equ14}). When $E$ lies 
within this range, we are within the `continuous band' of {\it extended 
states} (in the reciprocal space). This means that the density of states 
corresponding to Eq.~(\ref{equ13}) is non-zero at all energies lying within 
this range and therefore $\phi_{m,2} \ne 0$ irrespective of the choice of 
$\Delta$. Therefore, 
\begin{equation}
{S}_{21}g_{m,1}+{S}_{22}g_{m,2} \ne 0 
\label{equ17}
\end{equation}
for all $m$ in {\it dual space}. This implies that, 
\begin{equation}
\lim_{n \rightarrow \infty} [{S}_{21}f_{n,1}+{S}_{22}f_{n,2}] = 0 
\label{equ18}
\end{equation}
in {\it real space}. $S_{ij}$ are the elements of the matrix {$\mathbf S$}. 
If the average density of states corresponding to Eq.~(\ref{equ14}) is 
non-zero, then Eq.~(\ref{equ14}) tells us that $\lim_{m \rightarrow \infty} 
\phi_{m,1}=0$, as we have chosen $\Delta > \lambda$ \cite{soko85}. 
This, implies that, 
\begin{equation}
\lim_{m \rightarrow \infty} [{S}_{11}g_{m,1}+{S}_{12}g_{m,2}] = 0 
\label{equ19}
\end{equation}
in dual space, and, 
\begin{equation}
{S}_{11}f_{n,1}+{S}_{12}f_{n,2} \ne 0 
\label{equ20}
\end{equation}
for all values of $n$ in real space. Now, the density of states for an 
Aubry model is non-zero in the immediate neighborhood of $E=\gamma$ (the 
band-center)~\cite{soko85}. We can therefore definitely say from 
Eq.~(\ref{equ18}) and Eq.~(\ref{equ20}) that, both $f_{n,1}$ and $f_{n,2}$ 
will be non-zero for any 
\begin{figure}[ht]
{\centering
\resizebox*{8cm}{8cm}{\includegraphics{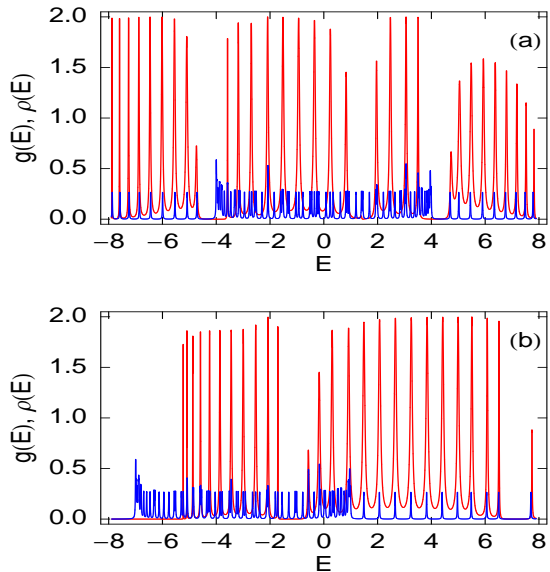}} \par}
\caption{$g$-$E$ (red color) and $\rho$-$E$ (blue color) curves for a ladder
of total number of rungs $60$. (a) $\gamma=0$ and (b) $\gamma=3$. Other 
parameters are, $Q=(1+\sqrt 5)/2$, $t_d=t_l=3$, $\epsilon_0=0$, $t_0=4$ and 
$\lambda=4$. We have chosen $c=e=h=1$.}
\label{cond}
\end{figure}
arbitrary value of $n$. This ensures that {\em all the states for the 
Aubry ladder will be extended at the center of the band}. However, 
interesting changes are observed as one looks away from the band-center. 
It is known~\cite{aubry79,soko85} that the density of states of 
an infinite one band Aubry model is highly fragmented. This means, we shall 
have zero values of the density of states (i.e., no state at all) scattered
throughout the spectrum. Whenever the density 
of states is zero, we shall have 
\begin{equation}
{S}_{11}f_{n,1}+{S}_{12}f_{n,2} =  0 
\label{equ21}
\end{equation}
for all $n$. Considering Eq.~(\ref{equ18}) and Eq.~(\ref{equ21}) together, 
it becomes quite clear that, both $f_{n,1}$ and $f_{n,2}$ have to be equal 
to zero as $n \rightarrow \infty$ in order 
that the Eqs.~(\ref{equ18}) and (\ref{equ21}) are simultaneously satisfied. 
This means that the eigenstates are localized away from the center of the 
band. We thus have extended wave functions at the band-center flanked by 
the exponentially localized states on either side for $-\lambda -\gamma < 
E < \lambda - \gamma$ and $\Delta > \lambda$.

It is to be noted that in the true quasiperiodic limit the spectrum of
an Aubry model exhibits more than one sub-bands separated by global gaps.
Within each sub-band one has a highly fragmented band structure with 
infinitesimal energy gaps (for an infinite system). So, in principle following
the argument given above, one should encounter an infinite number of mobility
edges. However, for realistic systems electron-electron interactions or
the lead-sample connection will broaden the energy levels and the 
infinitesimal gaps will not persists. Only mobility edges which reside in the
vicinity of the finite gaps separating the sub-bands will survive.
\vskip 0.05in
\noindent
{\em Case II:} $|E + \gamma| > \lambda$ and $\Delta > \lambda$
\vskip 0.05in
In this energy regime $E$ lies outside the band corresponding to the ordered 
system (Eq.~(\ref{equ13})), the corresponding density of states is zero (as 
there are no states at all). One then has, 
\begin{equation}
{S}_{21}f_{n,1}+{S}_{22}f_{n,2} =  0 
\label{equ22}
\end{equation}
for all $n$. If, on the other hand, the density of states corresponding to 
Eq.~(\ref{equ14}) is non-zero, then
\begin{equation}
{S}_{11}f_{n,1}+{S}_{12}f_{n,2} \ne  0 
\label{equ23}
\end{equation}
for all $n$. Therefore, from Eq.~(\ref{equ22}) and Eq.~(\ref{equ23}) we 
observe that, $f_{n,1}$ and $f_{n,2}$ both remain non-zero for all values 
of $n$. This means, the 
eigenstates are {\em extended} for these values of energy. Thus, for 
$\Delta > \lambda$, the eigenvalue spectrum for the Aubry ladder, in real 
\begin{figure}[ht]
{\centering
\resizebox*{8cm}{4cm}{\includegraphics{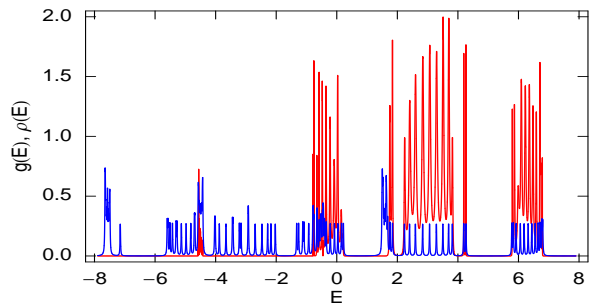}} \par}
\caption{$g$-$E$ (red color) and $\rho$-$E$ (blue color) curves for a general
AA-ladder of total number of rungs $60$. The parameters are, $\gamma=3$, 
$Q=(1+\sqrt 5)/2$, $t_d=1$, $t_l=2$, $\epsilon_0=0$, $t_0=4$ and 
$\lambda=4$. We have chosen $c=e=h=1$.}
\label{condgeneral}
\end{figure}
space, exhibits the existence of localized and extended states separated 
by {\em mobility edges} and a re-entrant {\em metal-insulator transition} 
is clearly visible. One can follow a similar chain of arguments to show 
that all states will be localized in real space for $\Delta < \lambda$. 
There are no mobility edges here.

For a more general choice of the Hamiltonian parameters (with $t_l \ne t_d$
and $\epsilon_{n,1} \ne \epsilon_{n,2}$), an analytical approach becomes 
difficult. We have numerically calculated the conductance of an Aubry ladder 
with various sets of parameters using a Green's function formalism. 
Though one has true quasiperiodicity only in an infinite system, for 
which we already have given an analytical proof for the existence of 
MI transition (for a special set of parameters), it is known that 
even finite systems grown following a quasiperiodic order are capable 
of exhibiting the localization effects~\cite{soko85,eco82,rolf90}.
For example, in finite laboratory-grown Fibonacci multilayers experimental 
evidence of localization of light has already been reported~\cite{suther}.
Therefore, though for an ideal infinite system several exotic spectral 
features may not be unlikely, in our cases of interest, as we have 
worked in the parameter regime where the single band AA-model has 
exponentially localized states only, expecting the localization 
fingerprints in a finite AA-ladder is quite legitimate. 
In the present calculation MI transition and mobility edges are found 
even when the simplification in the values of the hopping integrals are 
not made. 

To calculate the conductance, a finite Aubry ladder is attached to two 
semi-infinite one-dimensional metallic electrodes (Fig.~\ref{ladder}), 
described by the standard tight-binding Hamiltonian 
and parametrized by constant on-site potential $\epsilon_0$ and nearest 
neighbor hopping integral $t_0$. For low bias voltage and 
temperature, the conductance $g$ of the ladder is determined by the
Landauer conductance formula~\cite{datta} $g=(2e^2/h)T$ where the 
transmission probability $T$ is given by~\cite{datta} 
$T=Tr\left[\Gamma_S G_L^r \Gamma_D G_L^a\right]$. $\Gamma_S$ and $\Gamma_D$ 
correspond to the imaginary parts of the self-energies due to coupling of 
the ladder with the two 
electrodes and $G_L$ represents the Green's function of the ladder. All the 
informations about the ladder-to-electrodes coupling are included into these 
two self-energies as stated above and are described by the Newns-Anderson 
(NA) chemisorption theory~\cite{new,muj1}. 
For the sake of simplicity, here we have assumed that the entire voltage is 
dropped across the ladder-electrode interfaces and this assumption doesn't
significantly affect the qualitative aspects of the $g$-$E$ 
characteristics~\cite{datta,tian}.
In Fig.~\ref{cond} we present the behavior of the conductance for the cases 
when $\gamma=0$, and $t_l=t_d$, and the general case for a non-zero $\gamma$, 
and $t_l \ne t_d$ is shown in Fig.~\ref{condgeneral}. 
In every case the pictures of the density of states are superposed to show 
clearly that we have eigenstates existing in energy regimes for which the 
conductance is zero. This spikes of localized eigenstates and the transition 
from the conducting (high $g$) to non-conducting phase is clearly visible 
in each case.

Before we end, it should be pointed out that though the results presented 
in this communication are for zero temperature, they should be valid even
for finite temperatures ($\sim 300$ K) as the broadening of the energy 
levels of the ladder due to its coupling with the electrodes will be much 
larger than that of the thermal broadening~\cite{datta}. The 
inter ladder hopping $\gamma$ will shift the spectra
corresponding to Eqs.~(\ref{equ13}) and (\ref{equ14}) relative to each other,
thus making it possible, in principle, to tune the positions of the mobility
edges. This aspect may be utilized in designing a tailor made switching
device.

\end{document}